\documentclass[prl, twocolumn, superscriptaddress,floatfix]{revtex4}%

\usepackage{amsfonts}
\usepackage{amsmath}
\usepackage{amssymb}
\usepackage{graphicx}
\usepackage{epstopdf}
\usepackage{textcomp}
\usepackage[normalem]{ulem}
\usepackage{color}
\usepackage{gensymb}

\begin{document}

\preprint{}

\title{Anisotropic upper critical field of pristine and proton-irradiated single crystals of the magnetically ordered superconductor RbEuFe$_4$As$_4$}

\author{M. P. Smylie}
\email{matthew.smylie@hofstra.edu}
\affiliation{Materials Science Division, Argonne National Laboratory, 9700 S. Cass Ave., Lemont, Illinois 60439}
\affiliation{Department of Physics and Astronomy, Hofstra University, Hempstead, New York 11549}
\author{A. E. Koshelev}
\affiliation{Materials Science Division, Argonne National Laboratory, 9700 S. Cass Ave., Lemont, Illinois 60439}
\author{K. Willa}
\affiliation{Materials Science Division, Argonne National Laboratory, 9700 S. Cass Ave., Lemont, Illinois 60439}
\affiliation{Institute for Solid-State Physics, Karlsruhe Institute of Technology, 76021 Karlsruhe, Germany}
\author{R. Willa}
\affiliation{Materials Science Division, Argonne National Laboratory, 9700 S. Cass Ave., Lemont, Illinois 60439}
\affiliation{Institute for Theory of Condensed Matter, Karlsruhe Institute of Technology, 76131 Karlsruhe, Germany}
\author{W.-K. Kwok}
\affiliation{Materials Science Division, Argonne National Laboratory, 9700 S. Cass Ave., Lemont, Illinois 60439}
\author{J.-K. Bao}
\affiliation{Materials Science Division, Argonne National Laboratory, 9700 S. Cass Ave., Lemont, Illinois 60439}
\author{D. Y. Chung}
\affiliation{Materials Science Division, Argonne National Laboratory, 9700 S. Cass Ave., Lemont, Illinois 60439}
\author{M. G. Kanatzidis}
\affiliation{Materials Science Division, Argonne National Laboratory, 9700 S. Cass Ave., Lemont, Illinois 60439}
\affiliation{Department of Chemistry, Northwestern University, Evanston, Illinois, 60208, USA}
\author{J. Singleton}
\affiliation{National High Magnetic Field Laboratory, Los Alamos National Laboratory, MS-E536, Los Alamos, New Mexico 87545}
\author{F. F. Balakirev}
\affiliation{National High Magnetic Field Laboratory, Los Alamos National Laboratory, MS-E536, Los Alamos, New Mexico 87545}
\author{H. Hebbeker}
\affiliation{Department of Physics and Astronomy, Hofstra University, Hempstead, New York 11549}
\author{P. Niraula}
\affiliation{Department of Physics, Western Michigan University, Kalamazoo, Michigan 49008}
\author{E. Bokari}
\affiliation{Department of Physics, Western Michigan University, Kalamazoo, Michigan 49008}
\author{A. Kayani}
\affiliation{Department of Physics, Western Michigan University, Kalamazoo, Michigan 49008}
\author{U. Welp}
\affiliation{Materials Science Division, Argonne National Laboratory, 9700 S. Cass Ave., Lemont, Illinois 60439}

\begin{abstract}
We present a study of the upper critical field, $H_{c2}$, of pristine and proton-irradiated RbEuFe$_4$As$_4$ crystals in pulsed magnetic fields of up to 65 T.
The data for $H_{c2}$ reveal pronounced downwards curvature, particularly for the in-plane field orientation, and a superconducting anisotropy that decreases with decreasing temperature.
These features are indicative of Pauli paramagnetic limiting.
For the interpretation of these data, we use a model of a clean single-band superconductor with an open Fermi surface in the shape of a warped cylinder, which includes strong paramagnetic limiting.
Fits to the data reveal that the in-plane upper critical field is Pauli paramagnetic limited, while the out-of-plane upper critical field is orbitally limited and that the orbital and paramagnetic fields have opposite anisotropies.
A consequence of this particular combination is the unusual inversion of the anisotropy, $H_{c2}^{ab} < H_{c2}^c$, of the irradiated sample at temperatures below 10 K.
The fits also yield an in-plane Maki parameter, $\alpha_M^{110} \approx$ 2.6, exceeding the critical value for the formation of the Fulde-Ferrell-Larkin-Ovchinnikov state.
Nevertheless, the current measurements did not reveal direct evidence for the occurrence of this state.
\end{abstract}

\date{\today}

\maketitle

\section{I. INTRODUCTION}

The recent discovery \cite{Kawashima2016,Liu2016} of layered, Eu-containing compounds of composition AEuFe$_4$As$_4$ (A = Rb, Cs) has introduced new members to the family of materials that display superconductivity coexisting with three-dimensional long-range local moment magnetic order.
This has previously been seen in $R$Rh$_4$B$_4$ \cite{MapleBooks}, $R$Mo$_8$S$_8$ \cite{FischerMaple}, nickel borocarbides $R$Ni$_2$B$_2$C \cite{Canfield1998} and EuFe$_2$As$_2$-derived materials \cite{Zapf2017} ($R$ = Rare Earth).
The separation of the magnetic moments and of the superconducting electrons into different, essentially isolated sublattices as well as orbitally-selective exchange and pairing interactions are thought to enable the existence of superconductivity despite the high concentration of localized magnetic moments \cite{Eisaki1994,Jeevan2008,Sprau2017,Ren2009}.
The Eu$^{2+}$ ions carry a large, spin-only moment of 7$\mu_B$ which, at a temperature of 15 K, and enter a long-range ordered state, in which the moments display easy-plane anisotropy and order ferromagnetically within an Eu-layer \cite{Smylie2018,Albedah2018}.
Along the $c$-axis, these ferromagnetic layers have possibly a helical arrangement \cite{ZahirUnpublished}.
The superconducting transition occurs at $\sim$36.5 K \cite{Kawashima2016,Liu2016}.
Thus, the Europium-containing Fe-based superconducting materials are unique since they display simultaneously remarkably high magnetic ordering and superconducting transition temperatures, implying sizable magnetic exchange interactions in the presence of strong superconducting pairing. 

Previous measurements \cite{Smylie2018,KWilla-arXiv} of the magnetic and superconducting properties of single-crystal RbEuFe$_4$As$_4$ at comparatively low magnetic fields have revealed high slopes of the upper critical field indicative of large orbital upper critical fields and extreme type-II behavior with Ginzburg-Landau parameters of $\kappa_c \approx$ 70 and $\kappa_{ab} \approx$ 100 for magnetic fields applied perpendicular and parallel to the Fe$_2$As$_2$ layers, respectively.
These findings are in line with the behavior typically seen in optimally doped iron-based superconductors \cite{Zhang2011,Lei2012,Gurevich2010}.
Large orbital upper critical fields suggest that paramagnetic limiting is important in determining the low-temperature upper critical field, as has been proposed for various Fe-based superconductors \cite{Zhang2011,Lei2012,Gurevich2010,Fuchs2009,Kano2009,Khim2010,JJo2009,Altarawneh2008}.
However, in contrast to other groups of iron-based superconductors, RbEuFe$_4$As$_4$ and related so-called 1144-compounds are intrinsically hole-doped to $\approx$ 0.25 holes/Fe.
Approximately the same doping level is established in other Fe-based superconductors through chemical substitution to achieve optimum $T_c$ \cite{Johnston2010,Hirschfeld2011,Hosono2016}.
By avoiding chemical doping, long electron mean free paths and clean-limit superconductivity can be realized in RbEuFe$_4$As$_4$.
As a result, this material displays clean-limit extreme type-II superconductivity with strong paramagnetic limiting in a layered structure with a relatively low Fermi energy.
It is therefore a promising candidate for observing the Fulde-Ferrell-Larkin-Ovchinnikov (FFLO) state \cite{Gurevich2010,FuldeFerrell,LarkinOvchinnikov,Matsuda2007,Takada1970} at low temperatures and high magnetic fields.
Furthermore, a uniform exchange field originating from the completely polarized Eu-sublattice may further stabilize the FFLO state \cite{FuldeBooks}.

Here, we determine the upper critical field, $H_{c2}$, of pristine and proton-irradiated RbEuFe$_4$As$_4$ crystals using high-frequency susceptibility measurements in pulsed magnetic fields of up to 65 T applied parallel and perpendicular to the Fe$_2$As$_2$ layers.
The data for $H_{c2}$ reveal pronounced downwards curvature to values well below extrapolations based on  purely orbital Werthamer-Helfand-Hohenberg (WHH) theory \cite{WHH}, particularly for the in-plane field orientation, and a superconducting anisotropy that decreases with decreasing temperature.
These features are indicative of Pauli paramagnetic limiting.
A clean-limit theoretical description incorporating an open Fermi surface in the shape of a warped cylinder is developed.
Fits to the data suggest that the out-of-plane upper critical field is dominated by orbital limiting, whereas the in-plane upper critical field is dominated by Pauli paramagnetic limiting, and that the orbital and paramagnetic fields have opposite anisotropies.
A consequence of this particular combination is the unusual inversion of the anisotropy, $H_{c2}^{ab} < H_{c2}^c$, of the irradiated sample at temperatures below 10 K.
The fits also yield Maki parameters for which the in-plane orientation, $\alpha^{110} \approx$ 2.6, exceeds the critical value for the formation of the Fulde-Ferrell-Larkin-Ovchinnikov state.
However, no direct evidence for the occurrence of this state was observed in the current temperature and field range.

\section{II. EXPERIMENTAL METHODS}

Several high-quality, plate-like single crystals of RbEuFe$_4$As$_4$ with typical dimensions 500 $\mu$m x 500 $\mu$m x 80 $\mu$m were grown as described in Ref. \onlinecite{JinkeBao2018}.
The large faces of the crystals are perpendicular to the [001] direction and the long crystal sides are parallel to the [110] directions.
Three pristine samples were selected for pulsed field measurements (crystals \#1, \#2, \#3) and one sample was selected for proton irradiation prior to pulsed field measurements (crystal \#4).
All samples were pre-characterized using magnetization measurements performed in a 7 T Quantum Design MPMS with samples mounted on quartz rods with silicone grease.
Sample \#4 was irradiated along the $c$ axis with 5 MeV protons using the tandem Van de Graaff accelerator at Western Michigan University.
TRIM simulations \cite{Ziegler2015} for our irradiation geometry show that 5 MeV protons completely traverse the sample, creating a uniform density of defects that include point defects as well as collision cascades and clusters.
Following irradiation, the heat capacity of sample \#4 was measured using a membrane-based ac-nanocalorimeter in fields of up to 9 T \cite{Tagliati2012,KWilla-Inst}.
High-field susceptibility measurements were performed at the National High Magnetic Field Laboratory (NHFML) Pulsed Field Facility at Los Alamos National Laboratory (LANL) using the proximity diode oscillator (PDO) technique \cite{Nikolo2017,Altarawneh2009}.
The shift of the oscillator frequency with field and/or temperature is a measure of the degree of screening of magnetic flux in the sample which is either due to superconductivity or the normal-state skin depth; thus, the superconducting-normal transition is typically accompanied by a large shift in oscillator frequency.
The raw PDO frequencies are typically in the range 23 - 26 MHz.
Data shown in this paper represent these frequencies downshifted by a double heterodyne detection system to f $\approx$ 2 MHz \cite{Nikolo2017,Altarawneh2009}.
To minimize eddy currents and to maximize sample cooling in the pulsed-field measurements \cite{Nikolo2017}, samples were cut into smaller pieces $\sim$0.1 x 0.1 x 0.02 mm$^3$ and heat-sunk to a sapphire plate to which the calibrated Cernox thermometer is glued (see below).
Each sample was fixed, using silicone grease, to a 5-turn pancake PDO coil made from 50-gauge, high-purity Cu wire, also heat-sunk to the sapphire plate \cite{Altarawneh2009}.
Magnetic fields were provided by a 65 T pulsed magnet, with a rise time to peak field of about 9 ms, and a down-sweep time of about 90 ms; hence $dB/dt$ is much smaller as the field decreases \cite{Nikolo2017}.
Stable sample temperatures were obtained by placing the sample probe within a vacuum jacket inside a $^4$He bath; $^4$He exchange gas provided cooling, and temperatures were stabilized by varying the electrical power to a heater, disseminated over an extended length of the probe to avoid thermal gradients.

\section{III. RESULTS}

\begin{figure}
	\includegraphics[width=1\columnwidth]{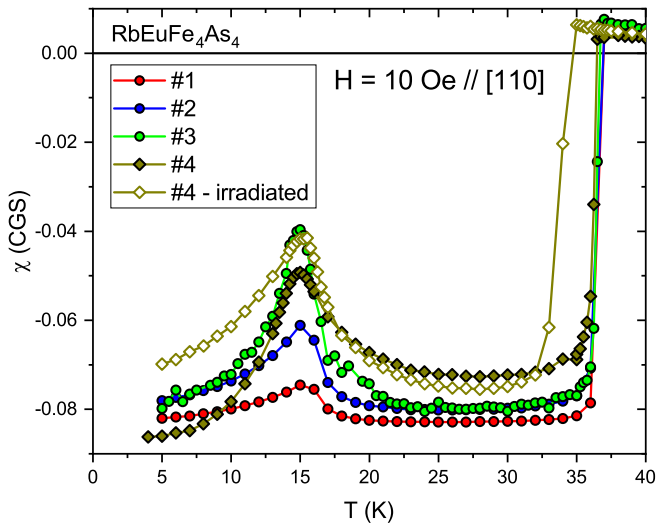}
	\caption{
		Temperature dependence of the susceptibility of the crystals used in this study.
		Data are taken in a field of 10 Oe applied along the [110] direction after zero-field cooling. 
	}
	\label{Fig1-xT}
\end{figure}

The magnetic and superconducting transitions of the crystals were characterized with zero-field cooled (ZFC) magnetization measurements.
Fig. \ref{Fig1-xT} shows ZFC data in a field of 10 Oe applied parallel to [110].
The pristine samples display a sharp diamagnetic transition at approximately 36.5 K, which is very reproducible from sample to sample.
Sample \#4 was irradiated with 5-MeV protons to a dose of 5$\times10^{16}$ p/cm$^2$.
This dose value was chosen based on previous experience with proton irradiation on optimally-doped Ba$_{0.6}$K$_{0.4}$Fe$_2$As$_2$ single crystals \cite{Kihlstrom2013}, which are thought to be electronically similar \cite{Teknowijoyo2018} to 1144-type materials.
Following the irradiation, there is a clear suppression of $T_c$ by approximately 2 K -- comparable to what is seen in proton-irradiated Ba$_{0.6}$K$_{0.4}$Fe$_2$As$_2$ -- likely originating from enhanced interband scattering \cite{Glatz2011}.
A cusp-like feature in the susceptibility near 15 K seen in all samples signals the magnetic ordering of the Eu moments.
Although the magnetic-ordering temperature is the same between samples, there is some sample-to-sample variability in the height and shape of the magnetic transition.
We attribute this to details of the sample shape and vortex pinning as the fields generated by the superconducting currents affect the way the Eu-moments magnetize \cite{VVV}.
In contrast to the onset of superconductivity, the magnetic transition temperature is essentially unchanged upon irradiation.

\begin{figure}
	\includegraphics[width=1\columnwidth]{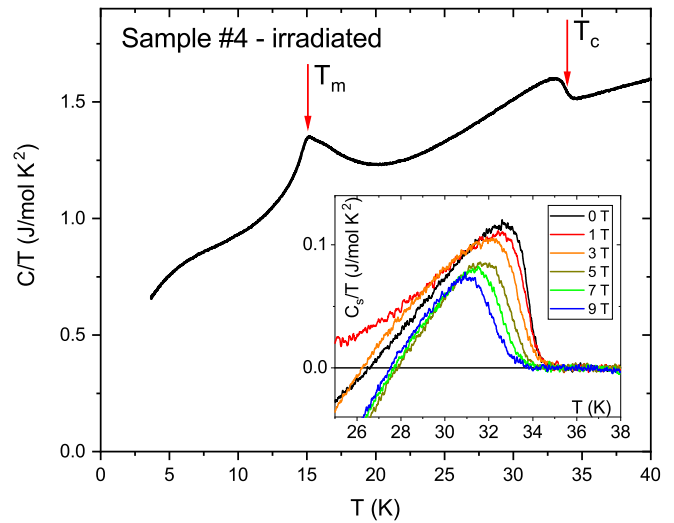}
	\caption{
		Temperature dependence of the zero-field heat capacity divided by temperature $C/T$ of the irradiated sample revealing clear signatures of the magnetic and superconducting transitions near 15 K and 34 K, respectively.
		The inset shows the evolution of the superconducting heat capacity, obtained by subtracting a linear extrapolation of the normal state signal, with magnetic fields applied along the $c$-axis.
	}
	\label{Fig2-CT}
\end{figure}

We also performed a calorimetric characterization on the irradiated sample shown in Fig. \ref{Fig2-CT}.
In a manner similar to pristine samples \cite{Smylie2018,KWilla-arXiv}, the magnetic transition appears as a clear, non-singular cusp whereas a mean-field like step signals the superconducting transition.
However, upon irradiation, the transition broadened.
The inset shows the evolution of the superconducting transition in magnetic fields of up to 9 T along the $c$-axis.
The unconventional sequence of the curves below the transition results from magnetic contributions to the heat capacity that in high fields superimpose onto the superconducting transition \cite{KWilla-arXiv}.
Using an entropy-conserving construction \cite{KWilla2018}, we deduce from this data set and from the corresponding $ab$-data upper critical field slopes of -8.8 T/K and -5.6 T/K for the $ab$ and $c$-directions, respectively.
The corresponding values for the pristine material are -7.1 T/K and -4.0 T/K \cite{Smylie2018,KWilla-arXiv}.
Upon irradiation the upper critical field slopes increase and the anisotropy is slightly reduced.
This might be expected as the superconducting coherence length decreases with increasing irradiation-induced electron scattering.

\begin{figure*}
	\includegraphics[width=2\columnwidth]{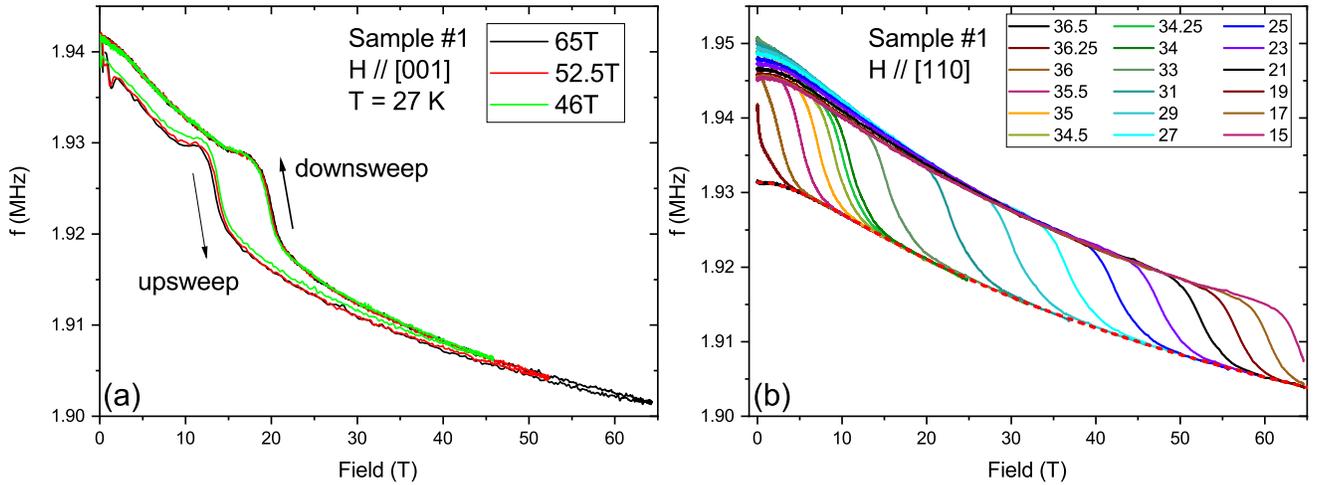}
	\caption{
		(a) Field dependence of the downshifted PDO frequency for a pristine RbEuFe$_4$As$_4$ crystal (sample \#1) measured for multiple pulse heights at T = 27 K with $H \parallel$ [001].
		There is hysteresis due to vortex motion (and to a lesser extent, eddy-current) heating on the up-sweeps.
		Whereas the transition in up-sweep data systematically shifts to lower fields with increasing pulse height.
		(b) Down-sweep field dependence of the downshifted PDO frequency at multiple temperatures with $H \parallel$ [110] on the same crystal.
		The collective background above the transition is well described by a 9th-order polynomial fit, indicated by the dashed red line.
	}
	\label{Fig3Layout}
\end{figure*}

Fig. \ref{Fig3Layout}(a) shows the field dependence of the downshifted PDO frequency at T = 27 K with $H \parallel$ [001] for crystal \#1 taken during three pulses with different peak fields.
The transition into the normal state is seen as a steep decrease of the oscillator frequency.
However, hysteresis is apparent between data recorded on the up-sweep and the down-sweep of the magnetic field.
Most of this is known \cite{Nikolo2017} to be caused by heating due to dissipative vortex motion in the mixed state during the rapid upsweep of the field.
Therefore, the sample is relatively hot as it exits the vortex state at $H_{c2}$.
This results in a shifted transition field, the position of which depends slightly on the field-pulse height; larger pulse heights give faster sweep rates, and hence less time for the heat to dissipate.
However, once the samples are in the normal state, there is known to be little heating due to the changing field; as mentioned above, sample sizes are kept (i) small to present very little cross-sectional area to the field (thereby minimizing eddy-current heating) and (ii) thin to provide a large surface area-to-volume ratio to maximize cooling.
In addition, rapid thermalization is assisted by thermal contact to the sapphire plate and by using a relatively high pressure of $^4$He exchange gas.
Finally, during the down-sweep, $dB/dt$ is significantly smaller than during the up-sweep, further reducing any residual eddy-current heating.
Hence, the sample is essentially in equilibrium with the sapphire plate (and the thermometer attached to it) when it enters the vortex state on the way down, leading to an accurate position for the transition.
Consequently, the down-sweep traces in Fig. \ref{Fig3Layout}(a) for different pulse heights essentially coincide (as do traces for samples of different sizes), and we use the down-sweep data for determining the phase diagrams. 

\begin{figure*}
	\includegraphics[width=2\columnwidth]{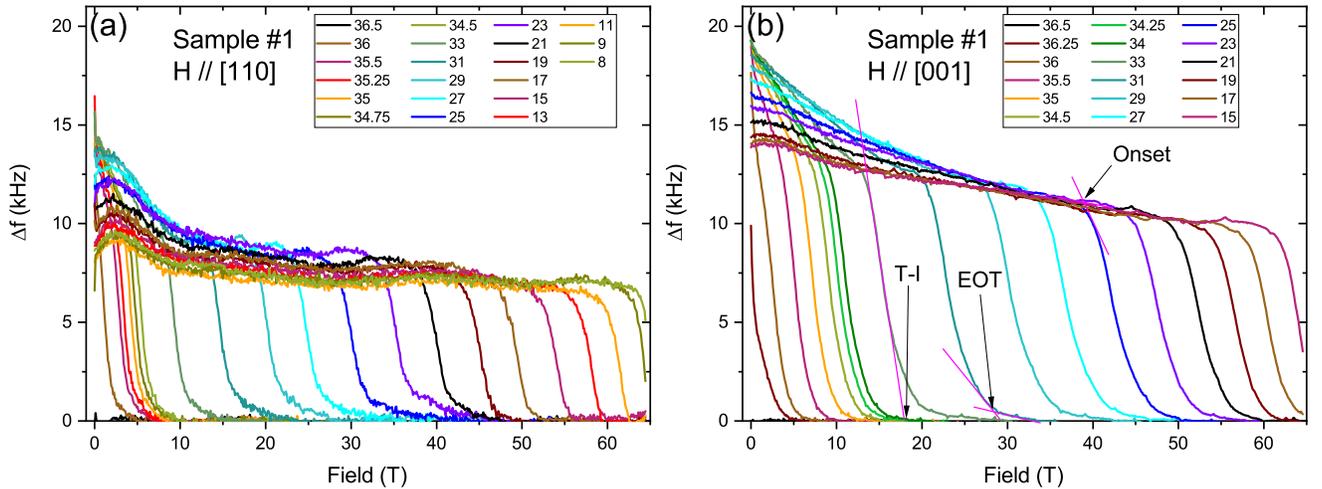}
	\caption{
		Frequency-shift PDO measurements at multiple temperatures of sample \#1 in pulse fields up to 65 T, with (a) $H \parallel$ [110] and (b) $H \parallel$ [001].
		The color scheme is identical for both figures.
		In panel (b), different transition criteria are indicated, including tangent-intercept (T-I), end-of-transition (EOT), and onset (Onset).
		For all field values, the transitions remain sharp and approximately parallel. 
	}
	\label{Fig4Layout}
\end{figure*}

The field dependence of the superconducting state, the magnetoresistance in the normal state of the sample, and the magnetoresistance of the pick-up coils \cite{Nikolo2017,Altarawneh2009} each contribute to an overall background signal onto which the superconducting transition is superimposed, see Fig. \ref{Fig3Layout}(a).
By offsetting the frequency, the normal state background signals at various temperatures can be collapsed onto a single smooth trace as shown in Fig. \ref{Fig3Layout}(b) (the down-sweep data, $H \parallel$ [110]).
A 9th-order polynomial can be fitted to the overall normal-state behavior above the transition for pulses at multiple temperatures; this is shown dashed in red in Fig. \ref{Fig3Layout}(b).
Subtracting this background for $H \parallel$ [110] and a corresponding curve for $H \parallel$ [001], the superconducting transitions and their evolution with temperature are clearly revealed as shown in Fig. \ref{Fig4Layout}(a) and (b) for crystal \#1; in the two panels the color schemes represent the same temperatures. 

\begin{figure}
	\includegraphics[width=1\columnwidth]{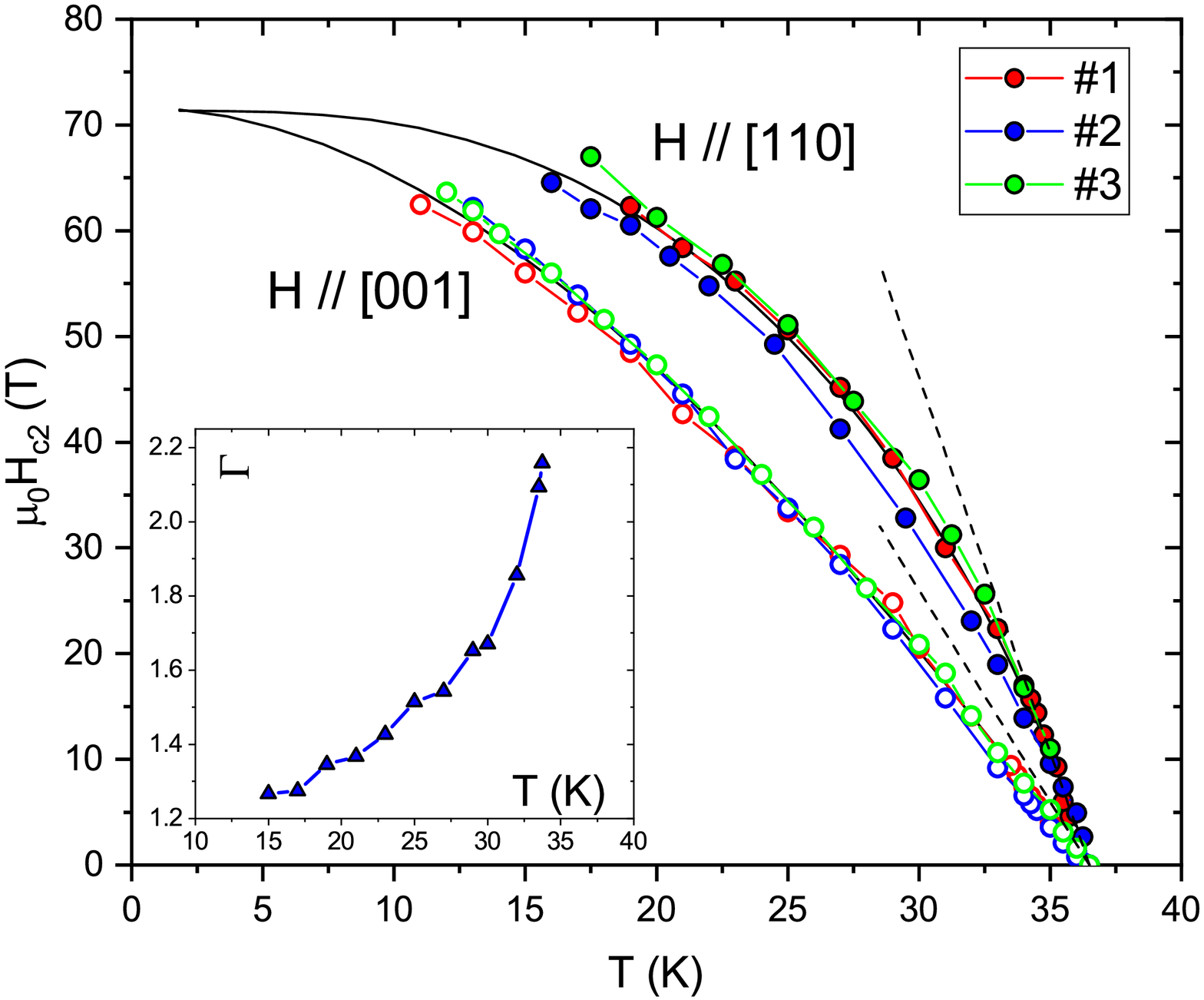}
	\caption{
		The upper critical field of the pristine crystals of RbEuFe$_4$As$_4$ with $H \parallel$ [110] (filled circles) and $H \parallel$ [001] (hollow circles).
		The dashed lines represent the high-temperature low-field Ginzburg-Landau (GL) slopes obtained from calorimetry [14].
		The solid lines are theoretical fits as described in the text.
		The inset shows the anisotropy ratio $\mathit{\Gamma}$ of crystal \#2 decreasing strongly with decreasing temperature.
		No obvious effect is observed in $H_{c2}$ when $T_c$ is suppressed below the magnetic ordering temperature (15 K) as may be expected since the Eu-moments are completely polarized.
	}
	\label{Fig5-Hc2}
\end{figure}

Data shown are characteristic of all 3 pristine samples measured.
On decreasing temperature (increasing field) the transitions remain sharp and parallel.
This behavior is consistent with the transitions seen in the magnetoresistance of 1144 materials \cite{Smylie2018,Meier2016}.
The effects due to vortex liquid phases and thermal fluctuations are noticeable but not as prevalent as, for instance, in cuprate superconductors \cite{Blatter1994}.
Nevertheless, there are various ways of defining the transition point as indicated in Fig. \ref{Fig4Layout}(b).
Here we adopt the 'end of transition' (EOT), corresponding approximately to the 90\%-criterion, commonly used for resistive transitions, since it yields at high temperatures a phase boundary that is in good agreement with the thermodynamic determination.
A similar trend has been observed in magnetoresistance and magnetization data on RbEuFe$_4$As$_4$ \cite{Smylie2018} and magnetoresistance and heat capacity data on CaKFe$_4$As$_4$ \cite{Meier2016}.
The $H_{c2}$-lines are shown in Fig. 5 for both field orientations.
Also included are the high-temperature upper critical field slopes obtained from heat capacity measurements \cite{KWilla-arXiv} on a companion crystal of those measured in pulsed fields.  
We can see that the low-temperature upper critical fields for both field orientations exceed 65T.
The temperature dependence of the in-plane $H_{c2}$ is characterized by a very pronounced downward curvature leading to decreasing anisotropy factor plotted in the inset.
The FFLO state would reveal itself in the data as an upward curvature of the upper critical field below temperatures of about $T_c/2$, and as additional features in the field dependence of various quantities such as heat capacity, thermal expansion, NMR spectra or magnetization which indicate the transition from the uniform to the FFLO phase \cite{Matsuda2007,Singleton2000,Cho2011,Zocco2013}.
Our results shown in Figs. \ref{Fig4Layout} and \ref{Fig5-Hc2} do not yield an unambiguous signature of the FFLO state in the current field and temperature range.

\begin{figure*}
	\includegraphics[width=2\columnwidth]{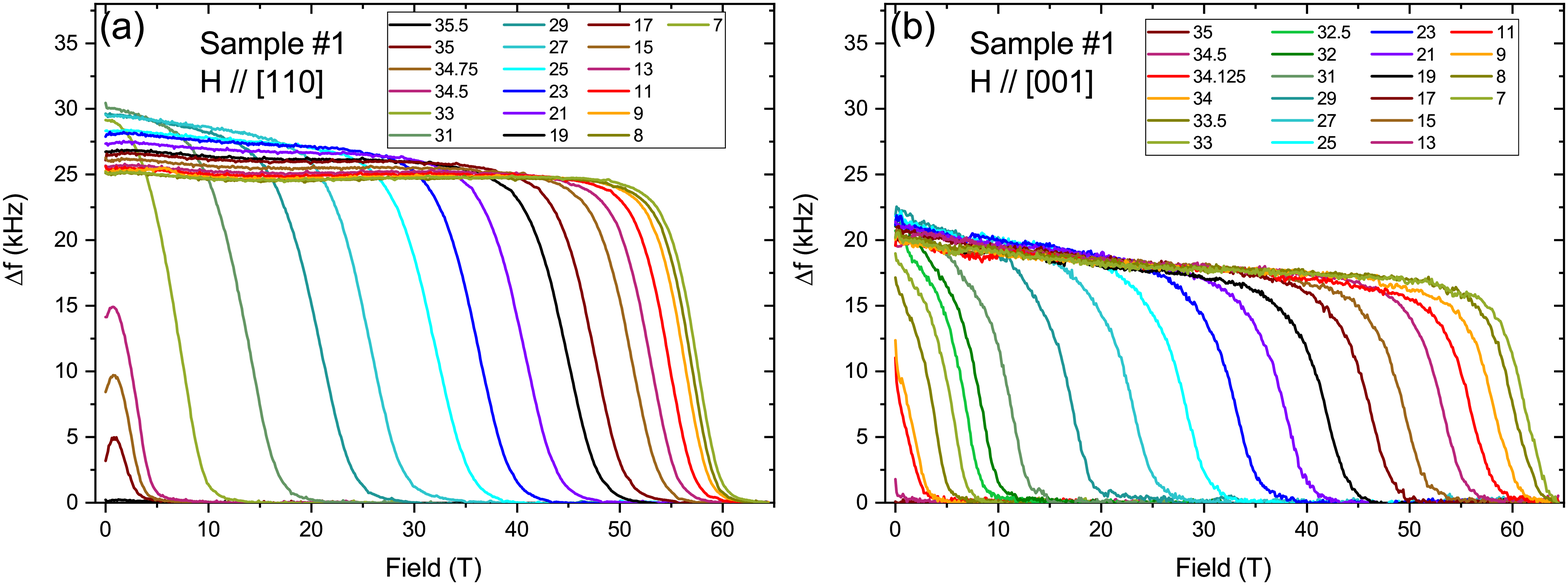}
	\caption{
		Frequency-shifted PDO data for measurements at multiple temperatures in the case of the irradiated sample \#4.
		Panel (a) shows $H \parallel$ [110] and panel (b) $H \parallel$ [001].
		For all field values, the transitions remain sharp and approximately parallel.
		It is immediately clear by examining the 25 K data (bright cyan) and the 7 K data (green-yellow) in both datasets, that at 25 K, $\mathit{\Gamma} = H_{c2}^{[110]}/Hc_{2}^{[001]} > 1$ whereas at 7 K, $\mathit{\Gamma} < 1$.
	}
	\label{Fig6Layout}
\end{figure*}

Pulsed-field measurements were performed on the irradiated sample \#4 in a manner identical to the pristine crystals.
Identical background subtraction techniques were applied, resulting in the data shown in Fig. \ref{Fig6Layout}(a) for $H \parallel$ [110] and in Fig. \ref{Fig6Layout}(b) for $H \parallel$ [001].
The transitions remain sharp and essentially parallel, indicating that the proton-induced disorder is uniform throughout the crystal.
In the two panels the color schemes represent the same temperatures; the anisotropy, while still obvious, is less than for the pristine samples, and at low temperatures, it clearly reverses. 

\begin{figure}
	\includegraphics[width=1\columnwidth]{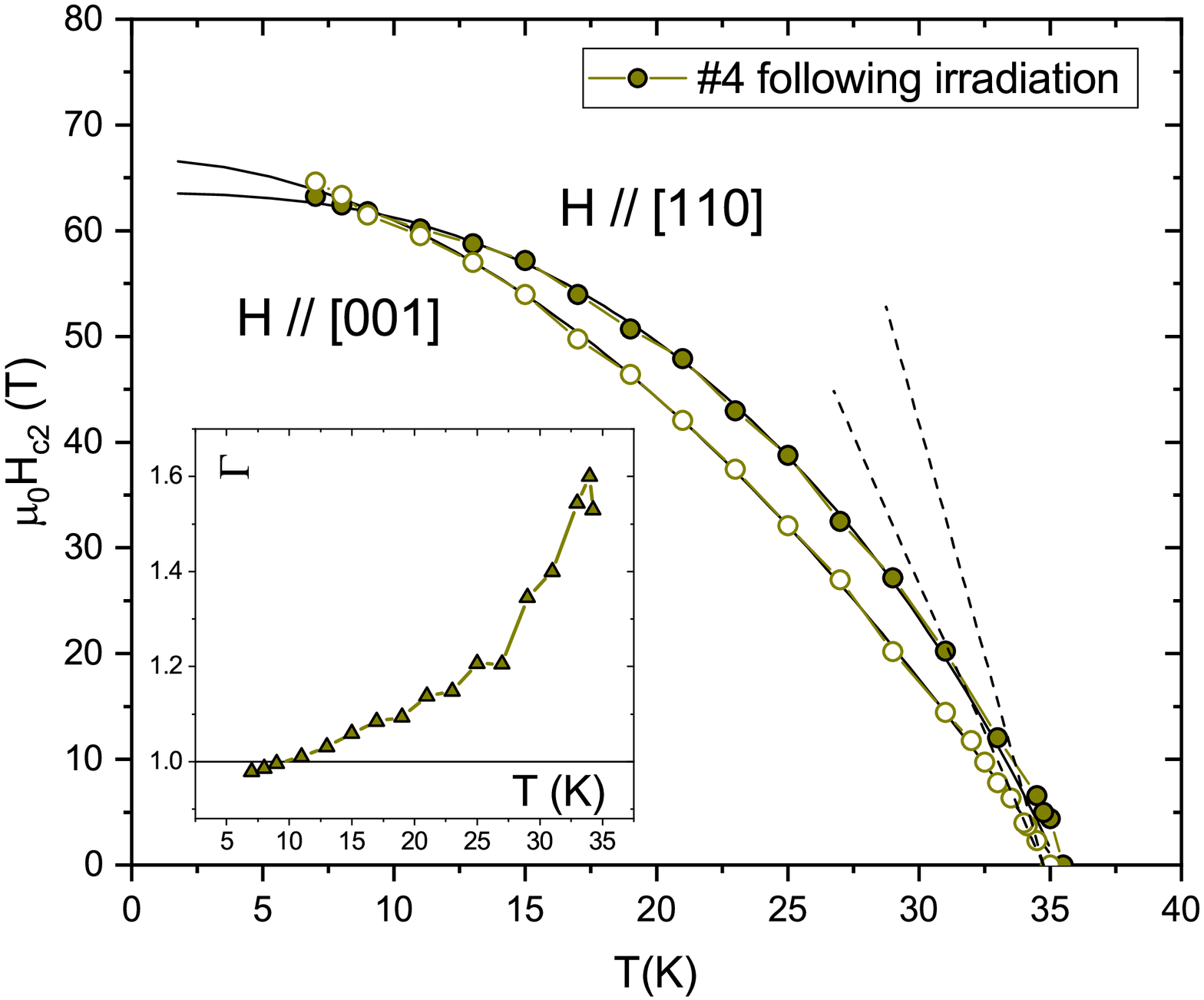}
	\caption{
		The upper critical field of a single crystal of RbEuFe$_4$As$_4$ which received a proton dose of 5$\times10^{16}$ p/cm$^2$ with $H \parallel$ [110] (filled symbols) and $H \parallel$ [001] (open symbols).
		The solid lines are fits as described in the text.
		The anisotropy ratio $\mathit{\Gamma} = H_{c2}^{[110]}/Hc_{2}^{[001]}$ as a function of temperature is shown in the inset.
		The anisotropy reverses at T $\approx$ 10 K.
	}
	\label{Fig7-Hc2Irr}
\end{figure}

The resulting phase boundaries are shown in Fig. \ref{Fig7-Hc2Irr} for both field orientations.
In comparison to the pristine samples (Fig. \ref{Fig5-Hc2}), the in-plane phase boundary of the irradiated crystal has clearly shifted to lower fields, while the out-of-plane upper critical field has decreased only slightly, consistent with the overall reduction of the anisotropy upon irradiation.
The inversion of the anisotropy, hinted at in the pristine samples, occurs near 10 K in the irradiated sample as is clearly seen in the temperature dependence of the anisotropy ratio $\mathit{\Gamma} = H_{c2}^{[110]}/Hc_{2}^{[001]}$ (inset of Fig. \ref{Fig7-Hc2Irr}).

\section{IV. THEORETICAL INTERPRETATION AND DISCUSSION}

In high magnetic fields, the superconducting state of a spin-singlet superconductor is suppressed according to two mechanisms.
Orbital pair breaking arises due to the Lorentz force acting on electrons paired with opposite momenta,
while paramagnetic pair breaking arises due to the Zeeman energy of the electrons paired with opposite spins. 

The zero-temperature orbital upper critical field, $H_{orb}$, of a single-band weak-coupling superconductor with ellipsoidal Fermi surface is given \cite{WHH} as $H_{orb} = -\left. 0.69~dH_{c2}/dT \right|_{T_c}$ in the dirty limit and $H_{orb} = -\left. 0.73~dH_{c2}/dT \right|_{T_c}$ in the clean limit.
The large values of $H_{c2}$ suggest that the superconducting coherence length is short, and the assumption of clean-limit behavior appears justified.
Indeed, we can estimate the electron mean free path, $l$, within a single-band Drude model that takes into account the low anisotropy as 
$l = \hbar\left( 3\pi^2n\sqrt{\epsilon} \right)^{1/3}/ ne^2\rho_n$, where $n$ = 1.25$\times10^{21}$ cm$^{-3}$ from polycrystalline Hall measurements \cite{Liu2016}, 
$\epsilon = m_{c}/m_{ab} = \left(\xi_{ab}/\xi_{c}\right)^2 \approx \left(1.4~\mathrm{nm}/0.92~\mathrm{nm}\right)^2 \approx 2.3$ 
is the ratio of effective masses, and $\rho_n \approx 20~\mu\Omega\cdot$cm is the in-plane resistivity at $T_c$ \cite{Smylie2018}.
The calculated value $l \approx 63$ nm qualifies the material as clean-limit, as $l \gg \xi_0$.
Currently, resistivity measurements on the irradiated sample are not available.
However, since in the pristine material $l$ is significantly larger than the coherence length, we assume that even after irradiation the sample is reasonably clean.
Then we estimate from the above relation the low-temperature orbital fields as 181 T ($H \parallel$ [110]) and 102 T ($H \parallel$ [001]) for the pristine crystals, and somewhat higher values 212 T ($H \parallel$ [110]) and 135 T ($H \parallel$ [001]) for the irradiated crystal.
As shown in Figs. \ref{Fig5-Hc2} and \ref{Fig7-Hc2Irr}, the experimental low-temperature values fall clearly short of these estimates indicating the importance of paramagnetic and/or possibly multi-band effects.
In absence of orbital effects, paramagnetic pair-breaking can be quantified by the paramagnetic limiting field $H_P$ (Chandrasekhar-Clogston limit) \cite{Chandrasekhar,Clogston}, the field at which the gain in Zeeman energy, $\mu_0 \chi_n H^2/2$, due to the normal-state spin susceptibility, $\chi_n = \mu_0 g^2 \mu_B^2 N(0)/2$, equals the superconducting condensation energy, $N(0)\Delta^2/2$, not considering any orbital effects: $\mu_0 H_P = \sqrt{2}\Delta/g\mu_B$.
Here, $\Delta$, $g$, $\mu_B$ and $N(0)$ are the superconducting gap, the electron $g$-factor ($g$ = 2 for free electrons), the Bohr magneton and the density of states per spin, respectively.
For a weak-coupling superconductor with $g$ = 2, one obtains the well-known result $\mu_0 H_P[T] = 1.85 T_c$ [K] \cite{Chandrasekhar,Clogston}.
The actual upper critical field, $H_{c2}$, of a spin-singlet superconductor is always smaller than both the paramagnetic limit $H_P$ and the orbital critical field $H_{orb}$, and is given in the dirty limit in the absence of spin-orbit scattering and for an ellipsoidal Fermi surface by \cite{Maki} $H_{c2} = H_{orb}/\sqrt{1+\alpha_M^2}$.
Here, $\alpha_M = \sqrt{2}H_{orb}/H_P$ is the Maki parameter measuring the relative role between orbital and paramagnetic pair breaking.
A similar simple relation does not exist for clean superconductors.
For most superconducting materials the paramagnetic limit is much larger than the orbital critical field ($\alpha_{M}\ll 1$), and the observed phase boundaries are well accounted for using orbital limiting only \cite{WHH}.
However, a universal feature of the iron-based superconductors are their extraordinarily large orbital critical fields \cite{Zhang2011,Lei2012,Gurevich2010} meaning that the paramagnetic effects may be essential.
For instance, a clean-limit theoretical description \cite{Gurevich2010} incorporating two ellipsoidal bands as well as paramagnetic and orbital limiting has been developed.
However, another nearly universal feature of iron-based superconductors is the geometry of their Fermi surfaces, which comprise several warped cylinders oriented along the $c$-axis. 
The warping induces substantial $c$-axis dispersion and the surprisingly low electronic anisotropy commonly seen in iron-based superconductors \cite{Yuan2009}.
In the case of RbEuFe$_4$As$_4$ no direct measurements of the electronic structure are currently available.
However, band structure calculations for this material \cite{Xu-arxiv} as well as ARPES \cite{Mou2016} and band structure calculations \cite{Lochner2017,Suetkin2018} on the non-magnetic sister compound CaKFe$_4$As$_4$ reveal the generic cylindrical Fermi surface topology for both electron and hole pockets.
This implies that for in-plane magnetic fields, a substantial fraction of electron orbits are open whereas for $c$-axis fields all orbits are closed.
In contrast, in theoretical models based on ellipsoidal Fermi surfaces, all orbits are closed \cite{AshcroftMermin}.

In general, the Fermi surface topology may have a strong influence on the orbital limit.
For example, for the case of layered superconductors with a very weak interlayer hopping integral, $t_\perp \ll k_B T_c$, the orbital upper critical field diverges at a certain temperature, i.e., a pure orbital effect does not destroy superconductivity at low temperatures \cite{Bulaevskii1973}.
In this case, the Maki parameter has no meaning.

A precise theoretical calculation of the upper critical fields requires detailed quantitative knowledge of the electronic band structure, Cooper-pairing interactions, and $g$-factors.
Such information is currently not available for RbEu1144.
For an approximate description of experimental data, we use the model of a single-band $s$-wave superconductor with a warped cylindrical Fermi surface described by the electronic spectrum $\epsilon(\mathbf{p}) = \frac{p_x^2+p_y^2}{2m} + 2t_\perp \mathrm{cos}\left(p_z d/\hbar\right) - \epsilon_F$ and anisotropic Zeeman energy $\mu_j H_j$. Here $m$ is the effective mass, $t_\perp$ is the interlayer hopping integral, $d$ is the interlayer period,  and $\mu_j = g_j \mu_B$ are the components of the electron's magnetic moments.
Such a simple model provides a reasonable description of a multiband material in the case when all bands have similar Fermi velocities.
Following the standard framework presented in Ref. \onlinecite{Kogan2012} based on the linearized Eilenberger equations, we obtain approximate closed-form equations for the upper critical field.
Here, we present only the summary of the results; the theoretical details will be published elsewhere \cite{Koshelev-Unpublished}.
The temperature-dependent in-plane upper critical field, $H_{c2,y}(T)$, is obtained as a solution of the equation
\begin{multline}
	-\mathrm{ln}~t = \int_0^\pi \frac{dq}{\pi} \int_0^\infty \frac{t~ds}{\mathrm{sinh}\left(ts\right)} \\ 
		\times\big[
			1 - \mathrm{cos}\left(\alpha_y h_y s\right) \mathrm{exp}\left(-u(q)h_y s^2\right) \\
			\times I_0 \left(u(q) h_y s^2\right) \mathrm{exp}\left(-2~\mathrm{sin}^2 q~h_y s^2\right)
		\big]
\end{multline}
where $u(q) = 1 - \frac{2t_\perp}{\epsilon_F} \mathrm{cos}~q$, $t = T/T_c$, $h_y = H_{c2,y}(T)/H_{y0}$, $H_{y0} = \frac{ 8\pi (k_BT_c)^2 \Phi_0 }{ \hbar^2 \bar{v}_x \bar{v}_z}$ (CGS units), $\alpha_y = \frac{ 8 \Phi_0 k_BT_c \mu_y }{ \hbar^2 \bar{v}_x \bar{v}_z}$, and $I_0(x)$ is the modified Bessel function of the first kind.
Furthermore, $\bar{v}_j = \sqrt{\langle v_j^2 \rangle}$ are the averaged Fermi velocities or, explicitly, $\bar{v}_{x,y} = v_F / \sqrt{2}$ with $v_F = \sqrt{2\epsilon_F / m}$ and $\bar{v}_z = \sqrt{2}t_\perp d/\hbar$.
We emphasize that this equation is approximate and it is not expected to work for very anisotropic materials for which $t_\perp < k_B T_c$.
The $c$-axis upper critical field, $H_{c2,z}(T)$, is determined by the exact equation
\begin{multline}
	-\mathrm{ln}~t = \int_0^\infty \frac{t~ds}{\mathrm{sinh}(ts)}
	\bigg[
		1 - \mathrm{cos}\left(\alpha_z h_z s\right) \mathrm{exp}\left(-2h_z s^2\right) \\
		\times I_0 \left(2 \frac{2t_\perp}{\epsilon_F} h_z s^2 \right)
	\bigg]
\end{multline}
with $h_z = H_{c2,z}(T)/H_{z0}$, $H_{z0} = \frac{16\pi (k_BT_c)^2 \Phi_0}{\hbar^2 v_F^2}$ , and $\alpha_z = \frac{16\Phi_0 k_BT_c \mu}{\hbar^2 v_F^2}$.
The Zeeman parameters $\alpha_j$ are related to but not identical with the Maki parameters.
The Maki parameters can be computed from the corresponding orbital fields, $H_{orb}$, which are solutions of the above equations at $\alpha_j = 0$ and $t\rightarrow 0$.

The fit parameters contained in this model are the ratio $t_\perp / \epsilon_F<0.5$ determining the warping of the Fermi surface, the orbital field scales, $H_{j0}$, and the Zeeman parameters $\alpha_j$ for each field direction.
The latter two parameters can be converted into the more conventional orbital critical fields and Maki parameters.
As shown in Fig. \ref{Fig5-Hc2}, the results for the upper critical field of the pristine crystals are fairly consistent between samples.
However, there are variations in the details of the shape of the $H_{c2}$-curve, which, in an unrestrained fit, lead to variations of the fit parameters.
In order to capture the general trends, we therefore fit an averaged $H_{c2}$-curve. 
We also fixed the warping parameter $t_\perp / \epsilon_F$ at two values, 0.2 and 0.3, as suggested by recent band structure calculations \cite{Xu-arxiv} leaving $H_{j0}$ and $\alpha_j$ as two independent fit parameters in Eqs. (1) and (2).
Fits to the data are included in Figs. \ref{Fig5-Hc2} and \ref{Fig7-Hc2Irr} as solid lines, and the resulting orbital and paramagnetic critical fields and Maki parameters computed from the fit parameters are listed in Table 1.
We note that for large values of the Maki parameters obtained for the in-plane orientation, the shape of the $H_{c2}(T)$ curve is actually weakly sensitive to the shape of Fermi surface, i.e., it very weakly depends on the warping parameter and is very close to the shape for the elliptical Fermi surface.

\begin{table}
\centering
 	\begin{tabular}{ | c | c | c | c | c | c | c | c |}
		\hline
		Sample & $t_\perp / \epsilon_F$ & $H_{orb}^{110}$ & $H_{orb}^{001}$ & $\alpha_M^{110}$ & $\alpha_M^{001}$ & $H_P^{110}$ & $H_P^{001}$ \\ \hline
		\#1,2,3 & 0.2 & 217 & 72 & 2.5 & $\textless0.1$ & 121 & $\textgreater1000$ \\ \hline
		\#1,2,3 & 0.3 & 223 & 75 & 2.6 & 0.34 & 121 & 310 \\ \hline
		\#4-irrad & 0.12 & 141 & 74 & 1.7 & 0.65 & 116 & 161 \\ \hline
		\#4-irrad & 0.2 & 147 & 79 & 1.8 & 0.84 & 116 & 162 \\ \hline
		\#4-irrad & 0.3 & 150 & 83 & 1.84 & 0.96 & 115 & 122 \\
		\hline
	\end{tabular}
\caption{
		Fit parameters of the superconducting phase diagram with both orbital and Pauli limiting for pristine crystals (\#1-\#3) and of crystal \#4 which has been irradiated with protons to a dose of 5$\times10^{16}$ p/cm$^2$.
		For the latter we also include the fitting parameters for a warping parameter of 0.12.
		}
\end{table}

A common trend emerging from all samples and from all fits is the observation that for the [110] direction the orbital critical field is clearly larger than the paramagnetic limit whereas for the [001] direction the inverse holds: the orbital critical field is smaller.
Thus, the upper critical field of RbEuFe$_4$As$_4$ is for the $c$-axis orbitally limited, while for the $ab$-plane paramagnetic limiting is significant.
Such a scenario has previously been described for highly-anisotropic organic superconductors \cite{Kogan2012} as well as other Fe-based superconductors \cite{Zhang2011,Gurevich2010}.
The shape of the $H_{c2}$-curves and the temperature dependence of the $H_{c2}$-anisotropy, $\mathit{\Gamma}$, is determined by the fine balance between these two effects.
In particular, while in the pristine samples $\mathit{\Gamma}$ appears to approach 1 at the lowest temperatures, for the irradiated sample the anisotropy inverts, $\mathit{\Gamma} < 1$, below T $\approx$ 10 K.
The fits indicate that this inversion results from a slight increase of the $c$-axis orbital critical field and a slight reduction of the [110]-direction paramagnetic limit upon irradiation.
The inversion of the anisotropy of $H_{c2}$ of anisotropic superconductors is not common.
Signatures of an inverted $H_{c2}$-anisotropy have been reported for Fe$_y$Se$_{1-x}$Te$_x$ \cite{Khim2010,Fang2010} and Fe$_y$S$_{1-x}$Te$_x$ \cite{Lei2010}.
Pronounced inversion of the anisotropy has also been observed in quasi 1-D K$_2$Cr$_3$As$_3$ \cite{Fedor2015}.
However, this material lacks inversion symmetry, and therefore an admixture of a spin-triplet component to the order parameter is expected \cite{Smidman2017} which would greatly affect paramagnetic limiting.

The Pauli paramagnetic limiting fields obtained from our fits are anisotropic with the $c$-axis value being clearly larger than the in-plane values, while both orientations significantly exceed the Chandrasekhar-Clogston limit of $\mu_0 H_P[T] = 1.85 T_c$[K].
Large values of $H_P$ have been reported before (see Ref.\ \onlinecite{Khim2011} for a compilation of experimental results).
We note that in addition to $T_c$, $H_P$ can depend on \cite{Fuchs2009} gap anisotropy, multi-band effects, strong coupling, and spin-orbit coupling.
For instance, strong-coupling effects induce an additional factor $\left(1 + \lambda\right)^n$, where $n$ = 0.5 or 1 \cite{Orlando1979,Schossmann1989} and $\lambda$ is the electron-boson coupling constant.
Furthermore, in the expression for the paramagnetic limiting field, $\mu_0 H_P = \sqrt{2}\Delta/g \mu_B$, the electron $g$-factor is typically assumed to be $g$ = 2.
However, due to orbital contributions and/or spin-orbit coupling the $g$-factor of band electrons in a metal can be substantially larger or smaller than the free-electron value \cite{MacDonald1982}.
In addition, the $g$-factor can be anisotropic.
For instance, the normal-state magnetic susceptibility of non-magnetic CaKFe$_4$As$_4$ is anisotropic \cite{Meier2016}. Assuming that this anisotropy is mostly determined by the anisotropy of g-factors, 
we  can estimate $g^{001}/g^{110} \approx \left(\chi_n^{001} / \chi_n^{110}\right)^{1/2} \approx 0.9$. 
While these susceptibility data are consistent with a $c$-axis paramagnetic limiting field that is larger than the $ab$-value, the observed size of the anisotropy of $H_P$ exceeds the expected value.
Thus, the identification of the mechanisms underlying the enhanced values of $H_P$ and its anisotropy will require the determination of the $g$-factor, for instance by conduction electron spin resonance measurements or de Haas-van Alphen \cite{MacDonald1982} measurements.

Multiband orbital effects \cite{Gurevich2010,Golubov2003} may affect the $H_{c2}$-curve.
A two-band model taking into account orbital limiting alone used in other Fe-based superconductors \cite{Baily2009,Gurevich2007}, however, is not compatible \cite{Khim2010} with the always-convex curvature seen in our results, and as such, multiband orbital effects may not be as important as Pauli limiting.
Indeed, the authors of Ref. \onlinecite{Meier2016}. 
note that a two-band and single-band models of the upper critical field of non-magnetic CaKFe$_4$As$_4$ fit the data equally well. 

In the limit of strong paramagnetic limiting, the FFLO state can arise at low temperatures and high magnetic fields.
This state is characterized by a real-space modulation of the superconducting order parameter either in amplitude or phase \cite{Gurevich2010,FuldeFerrell,LarkinOvchinnikov,Matsuda2007,Takada1970} such that the system energy is minimized under the constraints of a large Zeeman energy and superconducting condensation energy.
For an isotropic $s$-wave superconductor the critical value of the Maki parameter beyond which the FFLO state is stable is $\alpha_{M_c}$ = 1.8 \cite{Gruenberg1966}.
However, this critical value depends strongly on the Fermi surface structure.
For instance, recent calculations for a quasi 2-D system with a single Fermi surface sheet in the form of a warped cylinder yield $\alpha_{M_c}$ = 4.76 for $c$-axis fields \cite{KokWee}.
The most favorable geometry for the FFLO instability in layered materials arises for a magnetic field oriented along the layer direction.
This geometry has been investigated in detail for the case of very small interlayer hopping $t_\perp \ll k_B T_c$ \cite{Bulaevskii1973}.
The opposite case, more relevant for iron pnictides, was never investigated.
The FFLO instability may also be influenced by multiband structure.
In the case of elliptical Fermi surfaces, this problem has been investigated in Ref. \onlinecite{Gurevich2010}.
The FFLO state has been observed for the in-plane geometry in highly layered organic superconductors \cite{Singleton2000} as well as the Fe-based superconductors LiFeAs \cite{Cho2011}and KFe$_2$As$_2$ \cite{Zocco2013}. 

\section{V. CONCLUSIONS}

We have determined $H_{c2}$ of pristine and proton-irradiated RbEuFe$_4$As$_4$ crystals using high-frequency susceptibility measurements in pulsed magnetic fields up to 65 T for fields applied parallel and perpendicular to the Fe$_2$As$_2$ layers.
The $H_{c2}(T)$ curves fall well below extrapolations based on GL or WHH theory, particularly for $H \parallel$ $ab$, and reveal a superconducting anisotropy that decreases with decreasing temperatures.
Such features suggest Pauli paramagnetic limiting.
Fits to the data based on our clean-limit theoretical model which accommodates a warped cylindrical Fermi surface reveal that the in-plane upper critical field is indeed Pauli paramagnetic limited, while the out-of-plane upper critical field is orbitally limited resulting in an uncommon inversion of the anisotropy, $H_{c2}^{ab} < H_{c2}^c$, of the irradiated sample at temperatures below 10 K.
Our fits also yield a high Maki parameter $\alpha_{M}^{110} \approx 2.6$ for $H \parallel$ $ab$, which exceeds the theoretical critical value necessary for formation of the FFLO state.
However, our current measurements did not yield direct evidence for an FFLO state, and further measurements in higher fields will be necessary to fully probe the low-temperature superconducting state of this unique compound.

\begin{center}
\textbf{ACKNOWLEDGMENTS}
\end{center}
This work was supported by the U. S. Department of Energy, Office of Science, Basic Energy Sciences, Materials Sciences and Engineering Division. K. W. and R. W. acknowledge support from the Swiss National Science Foundation through the Postdoc Mobility program.  A portion of this work was performed at the National High Magnetic Field Laboratory, which is supported by National Science Foundation Cooperative Agreements No. DMR-1157490 and No. DMR-1644779, and the State of Florida, as well as the Strongly Correlated Magnets thrust of the DoE BES ``Science in 100 T'' program.

\bibliographystyle{apsrev4}


\end{document}